\documentstyle[11pt,amssymb,epsf]{article}

\def\dalemb#1#2{{\vbox{\hrule height .#2pt
        \hbox{\vrule width.#2pt height#1pt \kern#1pt
                \vrule width.#2pt}
        \hrule height.#2pt}}}

\def\0{{\sst{(0)}}}
\def\1{{\sst{(1)}}}
\def\2{{\sst{(2)}}}
\def\3{{\sst{(3)}}}
\def\4{{\sst{(4)}}}
\def\5{{\sst{(5)}}}
\def\6{{\sst{(6)}}}
\def\7{{\sst{(7)}}}
\def\8{{\sst{(8)}}}

\def\td{\tilde}
\def\wtd{\widetilde}

\def\nn{\nonumber} \def\bd{\begin{document}} \def\ed{\end{document}}
\def\ds{\documentstyle} \let\fr=\frac \let\bl=\bigl \let\br=\bigr
\let\Br=\Bigr \let\Bl=\Bigl 
\let\bm=\bibitem
\let\na=\nabla
\let\pa=\partial \let\ov=\overline 
\newcommand{\be}{\begin{equation}} 
\newcommand{\ee}{\end{equation}} 
\def\ba{\begin{array}}
\def\ea{\end{array}}
\def\ft#1#2{{\textstyle{{\scriptstyle #1}\over {\scriptstyle #2}}}}
\def\fft#1#2{{#1 \over #2}}
\def\del{\partial}
\def\sst#1{{\scriptscriptstyle #1}}
\def\oneone{\rlap 1\mkern4mu{\rm l}}
\def\ie{{\it i.e.\ }}
\def\etc{{\it etc.\ }}
\def\via{{\it via}}
\def\semi{{\ltimes}}
\def\cv{{\cal V}}
\def\str{{\rm str}}
\def\jm{{\rm j}}
\def\im{{\rm i}}

\def\cramp{\medmuskip = 2mu plus 1mu minus 2mu}
\def\cramper{\medmuskip = 2mu plus 1mu minus 2mu}
\def\crampest{\medmuskip = 1mu plus 1mu minus 1mu}
\def\uncramp{\medmuskip = 4mu plus 2mu minus 4mu}

\def\mapright#1{\smash{\mathop{-\!\!\!-\!\!\!-\!\!\!-\!\!\!-\!\!\!
             \longrightarrow}\limits^{#1}}}
\def\maprightt#1#2{\smash{\mathop{-\!\!\!-\!\!\!-\!\!\!-\!\!\!-\!\!\!
             \longrightarrow}\limits^{#1}_{#2}}}

\def\tX{{{\wtd X}}}
\newcommand{\ho}[1]{$\, ^{#1}$}
\newcommand{\hoch}[1]{$\, ^{#1}$}
\newcommand{\bea}{\begin{eqnarray}} 
\newcommand{\eea}{\end{eqnarray}} 
\newcommand{\ra}{\rightarrow}
\newcommand{\lra}{\longrightarrow}
\newcommand{\Lra}{\Leftrightarrow}
\newcommand{\ap}{\alpha^\prime}
\newcommand{\bp}{\tilde \beta^\prime}
\newcommand{\tr}{{\rm tr} }
\newcommand{\Tr}{{\rm Tr} } 
\newcommand{\NP}{Nucl. Phys. }
\newcommand{\tamphys}{\it Center for Theoretical Physics\\
Texas A\&M University, College Station, Texas 77843}
\newcommand{\upenn}{\it Department of Physics and Astronomy\\
University of Pennsylvania, Philadelphia, Pennsylvania 19104}

\newcommand{\auth}{M. Cveti\v{c}\hoch{\dagger1}, 
H. L\"u\hoch{\dagger1} and C.N. Pope\hoch{\ddagger2}
 }

\begin{document}
\begin{flushright}
\hfill{CTP TAMU-03/00}\\
\hfill{UPR-874-T}\\
\hfill{hep-th/0002054}\\
\hfill{February 2000}\\
\end{flushright}

\vspace{10pt}

\begin{center}
{ \large {\bf Localised Gravity  in the Singular Domain Wall Background? 
}}

\vspace{10pt}
\auth

\vspace{10pt}

{\hoch{\dagger}\upenn}

\vspace{10pt}
{\hoch{\ddagger}\tamphys}

\vspace{40pt}

\underline{ABSTRACT}
\end{center}

We study singular, supersymmetric domain-wall solutions supported by
the massive breathing mode scalars of, for example, sphere reductions
in M-theory or string theory.  The space-time on one side of such a
wall is asymptotic to the Cauchy horizon of the anti-de Sitter (AdS)
space-time.  However, on the other side there is a naked singularity.
The higher-dimensional embedding of these solutions has the novel
interpretation of (sphere compactified) brane configurations in the
domain ``inside the horizon region,'' with the singularity
corresponding to the sphere shrinking to zero volume.  The naked
singularity is the source of an infinite attractive gravitational
potential for the fluctuating modes.  Nevertheless, the spectrum is
bounded from below, continuous and positive definite, with the wave
functions suppressed in the region close to the singularity.  The
massless bound state is formally excluded due to the boundary
condition for the fluctuating mode wave functions at the naked
singularity.  However, a regularisation of the naked singularity, for
example by effects of the order of the inverse string scale, in turn
regularises the gravitational potential and allows for precisely one
(massless) bound-state spin-2 fluctuating mode.  We also contrast
spectra in these domain wall backgrounds with those of the domain
walls due to the massless modes of sphere reductions.

{\vfill\leftline{}\vfill
\vskip 5pt
\footnoterule
{\footnotesize \hoch{1} Research supported in part by DOE grant 
DE-FG02-95ER40893 \vskip -12pt} \vskip 14pt
{\footnotesize  \hoch{2} Research supported in part by DOE 
grant DE-FG03-95ER40917.\vskip  -12pt}}

\pagebreak
\setcounter{page}{1}
\section{Introduction}

The study of domain walls in gauged supergravity theories has
attracted much attention over the past year. (For a review on an
earlier work, see \cite{CvSo}.)  In particular, since these solutions
are asymptotic to anti-de Sitter (AdS) space-times, they are of
importance in the study of the renormalisation group flows \cite{GPPZ}
of strongly coupled gauge theories in the context of the AdS/CFT
correspondence.  Typically these configurations are singular in the
interior, and asymptotically approach the boundary of the AdS
space-times.

    On the other hand static AdS domain walls (in $D=5$), which are
asymptotic to AdS Cauchy horizons on either side of a non-singular
wall, localise gravity on the wall \cite{RaSuI,RaSuII} (in $D=4$), and
thus have important phenomenological implications.  It is therefore of
great importance to find embeddings of gravity-trapping domain walls
within a fundamental theory such as a compactified string or M-theory,
a proposal initiated by H. Verlinde \cite{Verlinde}. Of particular
interest are field theoretic embeddings of such configurations in
supergravity theories that arise as effective theories of string and
M-theory compactifications.  For example, scalar fields in the
(abelian) vector supermultiplets of $N=2$ gauged supergravity theories
provide natural candidates for realising AdS domain walls.  However,
as it turns out these walls are in general singular in the interior,
and on either side of the wall they are asymptotic to the boundary of
AdS, as found for the one-scalar case in \cite{BeCvI,KLS} (and proven
in \cite{KaLi,BeCvII} for the multi-scalar case).  The result is,
essentially, a consequence of the fact that the gauged supergravity
potential for scalars in vector supermultiplets has at most one
non-singular supersymmetric extremum per non-singular domain, and this
extremum is a {\it maximum}.  (For a special choice of parameters for
the gauged supergravity scalar potential, one side of the domain wall
may asymptotically approach a non-AdS (``dilatonic'') space-time
\cite{Gub}.)

    Recently, we investigated \cite{dw9912} domain-wall space-times
supported by scalar fields belonging to massive supermultiplets.  In
particular, the breathing mode $\phi$ which parameterises the volume
of the sphere, and which is a singlet under the isometry group of the
sphere, provides a consistent truncation to a single massive scalar
mode. Its potential has a supersymmetric AdS {\it minimum} (and not a
maximum as in the case of massless vector supermultiplets) at
$\phi=0$. This allows for hybrid domain-wall solutions where on one
side of the wall in the transverse direction $\rho<0$ (in a co-moving
coordinate), the space-time is asymptotic (as $\rho \to -\infty$) to
the AdS horizon.  The other side of the wall (the $\rho>0$ region)
allows for two possibilities:
\begin{itemize} 
\item
Branch I is a {\it singular} domain-wall solution with $\rho\to 0^+$
corresponding to a naked singularity at which $\phi \to \infty$ (zero
volume of the sphere) \cite{dw9912}.
 
\item
Branch II is a non-singular domain wall with, $\rho<0$ corresponding
to the dilatonic wall domain \cite{Cv}, and with $\rho \to \infty$
reaching the sphere decompactification ($\phi \to -\infty$).  The
higher-dimensional interpretation of this solution is a spherically
compactified $p$-brane (in the $D=5$ case, a D3-brane of Type IIB
string theory compactified on the five-sphere $S^5$) in the domain
that extends from the horizon to the asymptotically flat space-time.
Unfortunately, this type of domain wall cannot trap gravity
\cite{dw9912}.
\end{itemize}
 
    The purpose of this paper is to investigate the singular
domain-wall solutions of Branch I.  Owing to the naked singularity,
such a solution is clearly undesirable from the classical point of
view.  However, quantum mechanics may be more tolerant of
singularities, and we shall investigate the fluctuation spectrum in
this singular background.  Interestingly, the attractive gravitational
potential is insufficiently singular at the naked singularity, thus
rendering the spectrum bounded from below. In particular, the boundary
condition on wave functions at the singularity allows for a continuous
spectrum with positive energy, and the massless bound state is
formally excluded.  We also point out that a regularisation of the
naked singularity, for example by modifying the metric at distances of
the order of the inverse string scale, renders the gravitational
potential finite and allows for precisely one (massless) bound-state
of the lower-dimensional graviton.  These issues are addressed in
section 2.

In section 3 we contrast the spectrum in the background of such a
singular breathing-mode domain-wall to those in domain-wall
backgrounds supported by massless scalars of sphere reductions. These
latter domain walls, which have been previously studied in the
literature \cite{FGPW,BS,BSI,CGLP,BBS}, also involve examples of
singular solutions in the interior.  However, on the AdS side they are
asymptotic to the AdS boundary.  In spite of the naked singularity, the
fluctuation spectrum is well behaved.  The spectrum of the fluctuating
modes sheds light on the spectrum of strongly-coupled gauge theories
via the AdS/CFT correspondence.

We also elucidate the nature of the higher-dimensional embedding of
this configuration in Section 4; it describes a $p$-brane
configuration in the domain ``inside the horizon.''  We show that the
the singularity structure of the massive scalar domain wall solution
is exactly the same as the massless scalar domain wall associated with
distributed $p$-branes with negative tension ingredients.  This
suggests that the inside of the horizon of non-dilatonic $p$-branes
such as D3-branes or M-branes should be included in the discussion and
the resolution of these singularities may be analogous to the one
proposed in \cite{repul} for the singularity associated with negative
tension.

\section{Singular domain wall of the breathing mode}

     The scalar arising as the breathing mode in a Kaluza-Klein sphere
reduction is massive, and in general yields a potential that allows a
supersymmetric minimum with negative cosmological constant.  The
effective Lagrangian for the breathing-mode scalar and gravity is of
the form:
\be
{\cal L}_D = e\, R - \ft12 e\, (\del\phi)^2 - e\, V\,,
\ee
where the potential is given by \cite{bdlps}
\be
V=\ft12\, g^2\, \Big(\fft{1}{a_1^2}\, e^{a_1\phi} - \fft{1}{a_1 a_2}\,
e^{a_2\phi}\Big)\,.\label{scalarpot}
\ee
The (positive) constants $a_1$ and $a_2$ are given by
\be
a_1^2 = \fft{4}{k} + \fft{2(D-1)}{D-2}\,,\qquad a_1\, a_2=
\fft{2(D-1)}{D-2}\,,
\ee
where $k$ is a certain positive integer.  For $D=4$, 7 and 5, this
integer takes the value $k=1$.  These cases correspond to the $S^7$
and $S^4$ reductions of $D=11$ supergravity, and the $S^5$ reduction
of type IIB supergravity, respectively.  For $D=3$ the integer $k$ can
be equal to 1, 2 or 3.  The case $k=1$ has a four-dimensional origin
as an $S^1$ Scherk-Schwarz reduction of the Freedman-Schwarz model.
The cases with $k=2$ and $k=3$ corresponding to $S^3$ and $S^2$
reduction of six-dimensional and five-dimensional supergravity
theories.

The potential has one isolated AdS minimum $\Lambda\equiv V_{\rm min}$
at $\phi=0$, tends to zero for $\phi \to -\infty$ (sphere
decompactification) and tends to infinity for $\phi \to +\infty$ (zero
volume of the sphere).  The domain-wall solutions were obtained in
\cite{bdlps}. In \cite{dw9912}, the analytic solution in terms of a
co-moving coordinate frame (simply related to the conformally flat
metric) was derived, and the nature of the solutions was analysed in
detail.  They occur in two distinct branches, associated with two
disconnected space-time regions.  Here, we shall just describe the
(asymptotic) behaviour of the solutions. (The analytical details can
be found in \cite{dw9912}.)

    Using the co-moving coordinate $\rho$, the metric is of the form
\be
ds^2=e^{2A}\, dx^\mu\, dx_\mu + d\rho^2\,.
\ee 
In the first branch, the coordinate $\rho$ runs from $-\infty$ to 0,
with
\bea
e^{2A} \sim e^{2c \rho}\,, & {\rm for} &  \rho\rightarrow 
-\infty\,,\nn\\
e^{2A} \sim \rho^\gamma, \ \ \gamma= {\ft{2a_2}{(D-2)\, a_1}}\,, 
 & {\rm for} &  
\rho \rightarrow 0\,.
\label{sgam}\eea
In the second branch, the coordinate $\rho$ runs from $-\infty$ to
$+\infty$, with
\bea 
e^{2A} \sim e^{2c \rho}\,,  & {\rm for} & \rho\rightarrow
-\infty\,,\nn\\ 
e^{2A} \sim \rho^\gamma \,,\quad  
\gamma= {\ft{2a_1}{(D-2)\, a_2}}\,,  & {\rm for} & 
\rho \rightarrow +\infty\,.  
\eea 
Thus we see that in both solutions, as $\rho \rightarrow -\infty$, the
metric becomes AdS, with the constant $c$ is given by $\Lambda=-(D-1)
(D-2)\, c^2$. ( Note that the Ricci tensor approaches $R_{\mu\nu} =
-(D-1)\, c^2\, g_{\mu\nu}$.)  Clearly, the asymptotically AdS region
$\rho\to \infty$ corresponds to the AdS Cauchy horizon.  In the second
branch, the solution is free of singularities, whilst in the first
branch there is a naked singularity at $\rho=0$.  We shall concentrate
on this first branch, and study the singular domain wall.

  Close to  the singularity point at $\rho=0$, the values of
the $\gamma$  coefficient appearing in (\ref{sgam}) 
for the various cases are summarised in Table 1.
\bigskip

\begin{center}
\begin{tabular}{|c|c|c|c|c|}\hline 
      & $D=3$ &$D=4$ & $D=5$ & $D=7$\\ \hline
$k=1$ & $\ft12$ & $\ft27$ & $\ft15$ & $\ft18$ \\ \hline
$k=2$ & $\ft23$ &         &         &\\ \hline
$k=3$ & $\ft34$ &         &         &\\ \hline
\end{tabular}
\end{center}
\bigskip

\noindent{ Table 1: The values of the coefficient $\gamma$ in
(\ref{sgam}) for the singular domain-wall solutions with the massive
breathing mode, in various dimensions $D$.}
\bigskip

\subsection{Spectrum of fluctuating modes}

    It is of interest to examine the quantum fluctuations around the
backgrounds of the Branch-I solutions. In particular, the spectrum may
suffer from pathologies due to the singular nature of the metric.

The fluctuations of the $D$-dimensional graviton (in an
appropriate  gauge) are described by a  
minimally-coupled scalar field in this gravitational background
\cite{RaSuII}. The spectrum of these fluctuating
modes in turn elucidates the nature of the modes in $(D-1)$
dimensions, and in particular the possibility of trapping a
$(D-1)$-dimensional massless graviton at such a domain wall.  (For a
detailed derivation of the equation for the fluctuating modes in
arbitrary dimensions, see \cite{CEHS}).  The minimally-coupled scalar
field $\Phi$ obeys the wave equation
\be
\del_\mu(\sqrt{-g}\, g^{\mu\nu}\, \del_\nu \Phi)=0\,.
\ee
We make the Ansatz $\Phi = e^{ip\cdot x}\, \chi(z)$, where $m^2 =p\cdot
p$ determines the mass of the fluctuating mode.  It is helpful to cast
the wave equation into the Schr\"odinger form, which can be done by
first writing the metric in a manifestly conformally-flat frame, as
\be
ds^2 = e^{2A(z)}\, (dx^\mu\, dx_\mu + dz^2)\,,
\ee
by means of an appropriate coordinate transformation.  For the 
Branch-I solutions that we are interested in, the coordinate $z$ runs
from $-\infty$ to 0, and $A(z)$ has the following asymptotic behaviour:
\bea
e^{2A} \sim \fft{1}{c^2\, z^2}\,,  & {\rm for}& z\rightarrow
-\infty\,, \nn\\
e^{2A} \sim z^{\td\gamma}\,,\qquad
\td \gamma = \fft{2\gamma}{2-\gamma}\,, 
&{\rm for} & z\rightarrow 0\,.\label{conformal}
\eea

   Making the field redefinition $\chi = e^{-(D-2)A/2}\, \psi$, the
wave equation assumes the form
\be
(-\del^2 - V)\, \psi = m^2\, \psi\,,
\ee
with the Schr\"odinger potential given by
\be
V = \fft{D-2}{2}\,  A'' + \fft{(D-2)^2}{4} \, (A')^2\,.
\ee
The asymptotic behaviour of the potential is given by
\bea
V \sim \fft{D(D-2)}{4z^2}\,,  &{\rm for}& z
\rightarrow -\infty\,,\nn\\
V \sim \fft{c}{z^2}\,, &{\rm for}& z\rightarrow 0\,.
\eea
Thus we see that the potential near the singularity
approaches a  negative infinity, and there the value of 
the negative constant $c$ {\it is} important.   It is given by
\be
c=-\ft14 + \fft{1}{(k+2)^2} >-\ft14\,,
\ee
which is independent of the dimension $D$.  
The full form of the Schr\"odinger potential  $V$ is sketched in Figure 1. 

\begin{figure}[ht]
\leavevmode\centering
\epsfbox{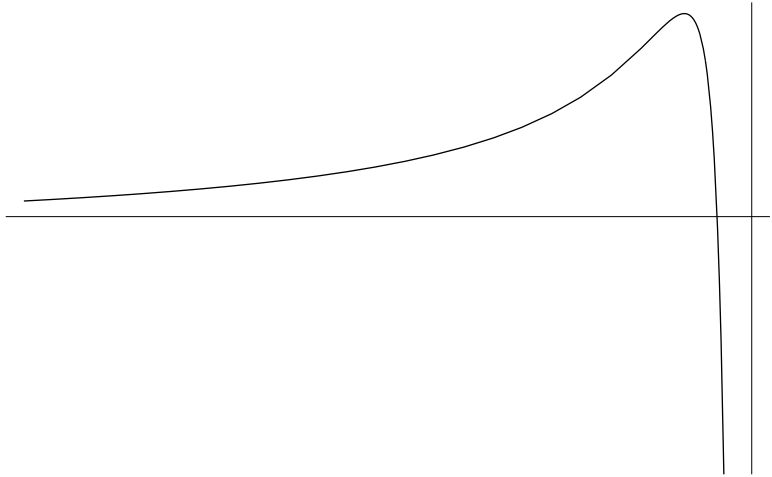}
\caption{A sketch of the Schr\"odinger potential $V(z)$ as a function of $z$.}
\end{figure}

    Since we have $c>-\ft14$, the spectrum is bounded from below.
Namely, the boundary condition at $\rho=0^-$ is $\Phi =0$, thus
disallowing solutions with $\Phi \ne 0$ at $\rho=0^-$. As a
consequence, the spectrum is continuous, with only positive energies
occurring.  Furthermore, the $m^2=0$ state has to be excluded, since
it corresponds to the solution $\phi=\,$ constant, which does not
vanish at $\rho=0$.  The ``bump'' resulting from the AdS space-time
causes the wave functions of the continuous spectrum to be be
suppressed in the interior region of the potential (close to the
singularity).  Thus although the background has a naked singularity,
the spectrum for minimally-coupled scalar fields is well behaved.
It does not, however, seem to be able to trap the massless
$(D-1)$-dimensional graviton.

          The Schr\"odinger potential for the branch-II solution is
quite different.  It runs smoothly from $z=-\infty$ to $z=+\infty$,
vanishing at both ends, with a single maximum at a certain finite value
of $z$.

\subsection{Regularising the metric near the naked singularity}

  It is on one hand encouraging that in spite of the naked singularity
the spectrum is well behaved, but on the other hand it is
disappointing that the massless mode is not bound (and is eliminated
by the boundary condition at the singularity).  However, this might
just be an artefact of working only at the level of the effective
supergravity theory.  Thus it might be that string-induced corrections
could ``regulate'' the metric near the naked singularity.

    We shall show in the next section that the singularity structures
of these massive-scalar domain-wall solutions are identical to those
of the massless-scalar domain walls.  The latter can be viewed in the
higher dimension as continuous distributions of D3-branes or M-branes
that include some negative-tension contributions.  A resolution of
singularities associated with negative-tension states was proposed in
\cite{repul}.  It is not inconceivable that a similar resolution can
be applied to our cases, since the singularity structure is the same.
As a consequence, the negative infinity of the Schr\"odinger potential
could be cut off at distances $z\sim M^{-1}_{string}$ and thus in fact
allow a zero-mass bound state after all.  In the absence of a direct
way of learning about these effects from string theory, here we
present a model where we modify (\ie regulate) the metric near the
singularity in terms of a plausible, albeit somewhat ad-hoc,
correction of order $M^{-1}_{string}$.  This does at least provide an
indication of the sort of modifications that can be expected once
stringy corrections are taken into account.
  
    Accordingly, near the singularity as $z\to 0^{-}$, and beyond
$z>0$, we shall modify the metric $A(z)$ so that it takes the form
\bea
e^{2A} = (|z|+M^{-1})^{\td \gamma}\,, & {\rm for}&  z\sim
\{-M^{-1}, 0^-\}\,, \\
e^{2A}  =  e^{-M' z}M^{-\td \gamma}\,, & {\rm for}&  z>0\,,
\eea
where $M'$ and $M$ are of the order of $M_{string}$.  The positive
coefficients $\td\gamma$ for the various examples are given by
(\ref{conformal}).  Clearly, taking $\{M,M'\}\to \infty$ corresponds
to the classical solution of the previous subsection.  (The metric is
continuous at $z=0$, but its higher derivatives are not.)

  Calculating the Schr\"odinger potential, we find that it now takes 
the following modified form:
\bea
V =  {c\over {(|z|+M^{-1})^2 }}\,, &  {\rm for} &   z\sim
\{-M^{-1}, 0^-\}\,, \\
V  =  \ft18 D(D-2)\, M'^2 \,,  & {\rm for}&  z>0\,.
\eea
Thus the negative infinity has been cut off with a step function, at
distances $z \sim -M^{-1}$. (See the sketch of the modified potential
in Figure 2).

\begin{figure}[ht]
\leavevmode\centering
\epsfbox{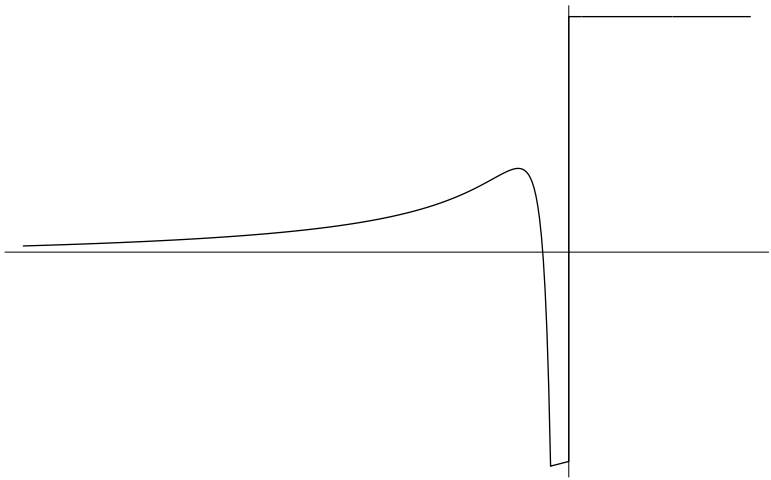}
\caption{A sketch of the regularised Schr\"odinger potential $V(z)$ as a
function of $z$. The negative infinity is cut-off at distances $z\sim 
-M^{-1}_{string}$  and at $z\ge 0$ a large barrier of order $M_{string}^2$
introduced.}
\end{figure}

  After the regularisation of the metric, the potential has been
rendered regular everywhere, and the Schr\"odinger equation can be
written as a supersymmetric quantum mechanical system, which allows
precisely one massless state (see for example \cite{DeWolf}): $\psi=
e^{(D-2)A(z)/2}$.  From the asymptotic behavior of the metric one sees
that in the ``regulated'' domain $z\ge 0^-$ the wave function falls
off exponentially fast, with a decay constant of order $M'^{-1}$.  On the
other hand the fast fall off in the AdS regime $z\to -\infty$ renders
the zero-mode renormalisable, and thus we have a {\it massless bound
state} trapped on the wall.

\section{Singular domain walls with ``massless'' scalars} 

        Recently a large class of AdS domain-wall solutions,
associated with diagonal symmetric potentials of lower-dimensional
gauged supergravities, were obtained
\cite{BS,FGPW,BSI,CGLP,BBS}. These solutions generally interpolate
between a boundary AdS and a naked singularity, and thus it is of
interest to revisit the properties of the fluctuation spectra here,
and to contrast them with those of the domain walls supported by the
massive breathing mode.  The metrics for the solutions supported by
the massless scalars are given by
\cite{CGLP}:
\bea
ds_{D}^2 &=& (gr)^{\ft{4}{D-3}}\, (\prod_i H_i)^{\ft12 -\ft2{N}}\,
dx^\mu\, dx_\mu + (\prod_i H_i)^{-\fft2{N}}\, \fft{dr^2}{g^2 r^2}\,,\nn\\
H_i&=& 1 +\fft{\ell_i^2}{r^2}\,,\qquad N=\fft{4(D-2)}{D-3}\,,
\eea
where $D=4, 5$ and 7, and for these dimensions we have $N=8$, 6 and 5
respectively.  The coordinate $r$ runs from zero to infinity.  At
$r=\infty$ we have $H_i=1$, and the metric describes AdS space-time in
horospherical coordinates.  Note that the AdS space-time is a maximum
of the scalar potential.

    As $r$ approaches zero, the metric behaviour depends on the number
of non-vanishing constants $\ell_i$. If $n <N$ of them are non-vanishing,
the metric approaches
\bea
ds_D^2 &\sim& \rho^{\gamma}\, dx^\mu\, dx_\mu + d\rho^2\,,\nn\\
\gamma &=& \fft{2(N-n)}{n(D-3)}\,,\qquad
\rho=r^{\ft{2n}{N}}\,.
\eea
The values of the constant $\gamma$ for the various AdS domain walls are
summarised in Table 2.  
\bigskip

\begin{center}
\begin{tabular}{|c|c|c|c|}\hline 
      & $D=4$ & $D=5$ & $D=7$\\ \hline
$n=1$ & 14    & 5     & 2    \\ \hline
$n=2$ & 6     & 2     & $\ft34$ \\ \hline
$n=3$ & $\ft{10}3$ &1  & $\ft13$ \\ \hline
$n=4$ & 2     &$\ft12$&$\ft18$ \\ \hline
$n=5$ & $\ft65$ & $\ft15$ & \\ \hline
$n=6$ &$\ft23$& &\\ \hline
$n=7$ &$\ft27$& &\\ \hline
\end{tabular}
\end{center}
\bigskip

\noindent{Table 2: The values of $\gamma$ coefficients
for the various domain walls supported by scalars of 
massless supermultiplets.}
\bigskip

        For all the cases, the metric has a power-law curvature
singularity $\sim 1/\rho^2$ at $\rho=0$.  For $\gamma \ge 2 $, the
singularity is marginal, in the sense that $\rho=0$ is also an
horizon, whilst for $\gamma < 2$ the singularity at $\rho=0$ is
naked.\footnote{There is an event horizon at $\rho=0$ if geodesics
originating at some finite and non-zero value of $\rho$ take an
infinite coordinate time to reach $\rho=0$.  Thus there is an horizon
when $\gamma\ge2$.}  These solutions, when oxidised back to $D=11$ or
$D=10$, become ellipsoidal distributions of M-branes or D3-branes.
One can in general argue that the naked singularities of these
lower-dimensional solutions are therefore artefacts of the dimensional
reduction.  However, in the case of the solutions with $n=N-1$, the
distributions involve negative-tension distributions \cite{FGPW,CGLP}
of the M-branes or D3-branes, which clearly also have naked
singularities in the higher dimension.  It is interesting to note that
for the cases associated negative tension distributions, the values
of $\gamma$ for $D=4$, 5 and 7 are precisely the same as the ones for
the massive-scalar breathing mode potentials given in Table 1.

    It is therefore worthwhile to investigate whether quantum
fluctuations around these backgrounds suffer from pathologies
associated with the naked singularities. The minimally-coupled scalar
$\Phi$ in these asymptotically AdS geometries is of special interest
(in the dual gauge theory it corresponds to the operator that couples
to the kinetic energy of the gauge field strength $F^2$).  The
spectrum for the wave equations of these background was studied in
detail in various publications, and no pathologies in the spectrum
were encountered in any of these cases \cite{CGLP,BBS}.  It is
straightforward to show that the wave equation near the singularity
$\rho=0$ is then of the form
\bea
&&\Big(-\del_z^2 - \fft{C_n}{(z - z_*)^2}\Big)\, \psi = m^2\, 
\psi\,,\nn\\
&& C_n=-\ft14 + \fft{(N-n-2)^2}{(N-n-4)^2}\,. 
\eea
We see that the coefficients always satisfy the bound $C_n \ge
-\ft14$, which is essential in order for the energies of all the
states to be bounded below.

    It is of interest to investigate the relation between the
structure of the spectrum and the nature of the curvature singularity
of the background.  It has been shown that the spectrum is discrete in
the cases $N-n=1$, 2 or 3.  For all of these values, the background
suffers from a naked singularity.  In fact, we can show that the
spectrum is not only discrete, but also positive definite.  This is
because the coordinate $r$ runs from $r=0$ to $r=\infty$, and hence
the wave function $\Phi$ has to satisfy the boundary conditions
$\Phi(0)=0$ and $\Phi(\infty)$ finite.  It is straightforward to see
that solutions with non-positive $m^2$ do not satisfy this condition.
In particular, the massless solution, {\it i.e.} $\Phi=1$, is excluded
by the boundary conditions, since it does not vanish at $r=0$.
Indeed, the case of $D=5$ and $n=4$ can be solved exactly, and the
spectrum can be seen to comprise only positive values of $m^2$.

        For $N-n\ge 4$, the singularity is marginally clothed by an
event horizon, and the spectrum is continuous, (with a mass gap when
$N-n=4$).  As we discussed earlier, the coordinate $\rho$ terminates
at $\rho=0$ in all the cases, owing to the singularity, and it follows
that the zero-mass solution has to be excluded from the spectrum since
$\Phi(0)$ must vanish.

      The absence of the $m^2=0$ states is consistent from the AdS/CFT
viewpoint, since one does not expect gravity to be localised on the
boundary of the AdS. Even in the case of a metric that is regulated at
$\rho \to 0$, the zero mass state mode is {\it excluded}; this mode is
of the form $\psi=e^{(D-2)A/2}$ (as obtained from the supersymmetric
quantum mechanical analysis).  However, this mode violates the
boundary condition that $\psi=0$ at the boundary of AdS.

\section{Higher dimensional interpretation}

      The higher-dimensional interpretations of the domain-wall
solutions supported by the massive breathing modes were given in
\cite{bdlps}.  They correspond to isotropic non-dilatonic branes in
the higher dimensions, {\it e.g.,} M-branes, D3-branes and self-dual
strings, \etc It is straightforward to see that the Branch-II solution
corresponds to the region of the $p$-brane that interpolates between
Minkowski space-time and the AdS throat.  The Branch-I solution, on
the other hand, corresponds to the region between the singularity (at
zero volume of the sphere) and the horizon, as we shall demonstrate
below.

     It was shown in \cite{garygary} that the D3-brane and M5-brane
admit maximal analytic extensions that do not have any singularities.
It is of interest therefore to examine the oxidation of the
massive-scalar domain wall to ten or eleven dimensions in more detail,
which we do by retracing the step of Kaluza-Klein reduction on the
sphere.  After doing this, the metric becomes \cite{bdlps}
\bea
ds_{\hat D}^2 &=& (\varepsilon\, H)^{2/({\hat D}-d-1)}\, 
dx^\mu\, dx_\mu +
          H^{-2} d\rho^2 + \rho^2\, d\Omega_d^2\,,\nn\\
H &=& 1 - \fft{Q}{\rho^{d-1}}\,,\label{oxidised}
\eea
in terms of an appropriately-defined Schwarzschild-type coordinate
$\rho$, where the constant $\varepsilon$ is $-1$ for the Branch-I
solution, and $+1$ for the Branch-II solution.  ($\hat D$ is the
oxidation endpoint dimension.) Branch-I solution therefore provides a
novel extention of $p$-brane solution into the interior of the
horizon.

     As was discussed in \cite{garygary}, when $\varepsilon=+1$ the
exterior region of a metric such as (\ref{oxidised}) (\ie the region
$\rho>Q^{1/(d-1)}$ outside the horizon) can be smoothly extrapolated
through the horizon at $\rho=Q^{1/(d-1)}$ and out into another
asymptotic region, thereby avoiding the curvature singularity at
$\rho=0$ altogether, provided that ${\hat D}-d-1$ is even.  This can
be seen by defining a new radial coordinate $w$,
\be
w= H^{\fft{1}{{\hat D}-d-1}}\,,
\ee
in terms of which the metric (\ref{oxidised}) becomes
\be
ds_{\hat D}^2 = w^2\, dx^\mu\, dx_\mu + \kappa^2\, 
\Big( 1-w^{{\hat D}-d-1}
\Big)^{-\fft{2d}{d-1}}\, \fft{dw^2}{w^2} + Q^{\fft2{d-1}}\, 
 \Big( 1-w^{{\hat D}-d-1} \Big)^{-\fft{2}{d-1}}\, d\Omega_d^2\,,
\ee
where $\kappa^2= ({\hat D}-d-1)^2\, Q^{\fft{2}{d-1}}/(d-1)^2$.  Since
$\rho$ is an analytic function of $w$ on the horizon at $w=0$, one can
analytically extend the metric to negative values of $w$, and so it is
regular on the horizon.  Furthermore, when ${\hat D}-d-1$ is even, the
metric is invariant under $w\longrightarrow -w$, and so the extension
to negative $w$ is isometric to the original region with positive $w$
\cite{garygary}.  Thus for the $D=7$ Branch-II domain wall oxidised on
$S^4$ to the M5-brane in $\hat D=11$ (for which we have $\hat D-d-1 =
11- 4-1=6$), and the $D=5$ Branch-II domain wall oxidised on $S^5$ to
the D3-brane in $\hat D=10$ (for which $\hat D-d-1=10-5-1=4$), the
metrics are completely non-singular.  On the other hand the $D=4$
Branch-II domain wall oxidises on $S^7$ to give an M2-brane in $\hat
D=11$, for which the singularity cannot be evaded by the
$w\longrightarrow -w$ reflection, since $\hat D-d-1=11-7-1=3$.

     In all cases the Branch-I solutions oxidise to the interior
regions of the M-branes or D3-brane, and so the singularities remain
in the higher dimension.  The fact that the Branch-I solutions map
into the interior regions of the higher-dimensional $p$-branes
demonstrates that these interior regions do have a r\^ole to play,
even in those cases where they are excised in the maximal analytic
extension.  From the lower-dimensional point of view, these naked
singularities are no worse than the ones occurring in the domain-wall
solutions associated with the Coulomb branches of the corresponding
superconformal field theories on the AdS boundaries, as observed in
section 3.

    The cases $\gamma=\ft27$, $\ft15$ and $\ft18$ can occur both from
the massless-scalar potential and from the massive-scalar potential.
For each dimension, these massless-scalar and massive-scalar solutions
have in common that their higher-dimensional origins both involve
naked singularities.  In the massless-scalar case, the singularity is
due to the negative tension of the distributed branes; in the
massive-scalar case, the singularity is the one that is inside the
horizon of the non-dilatonic $p$-brane.  However in spite of the
singularities, the minimally-coupled scalar field spectra are all
well-behaved.  This leads to an interesting question as to whether the
interior region should be included in the discussion, even in the
cases such as the M5-brane and D3-brane where it is normally excluded
by making the maximal analytic extension.  A resolution of the
singularity arising from a negative tension was proposed in
\cite{repul}.  A similar resolution may be applicable for the
massive-scalar domain-wall solution too, since the singularity
structure is the same.

\section*{Acknowledgements} 

We should like to thank Klaus Behrndt, Michael Cohen, Gary Gibbons,
Lisa Randall and Finn Larsen for useful conversations. M.C. would like
to thank Caltech Theory Group for hospitality where part of this work
was done.

\end{document}